\documentstyle[12pt,psfig,aaspp]{article}
\def\vg{\vec g}
\def\grad{\vec\nabla}
\def\vgx{\vec g_{ex}}
\begin{document}

\title{The modified dynamics is conducive to  galactic warp formation}
\author{ Rafael Brada and Mordehai Milgrom }
\affil{ Condensed Matter Department, Weizmann Institute of Science, Rehovot
 Israel}

\begin{abstract}

There is an effect in the modified dynamics (MOND)
 that is conducive to formation of warps. Because of the nonlinearity of the theory
 the internal dynamics of a galaxy is affected by a perturber
over and above possible tidal effects. For example, a relatively
distant and light companion or the mean influence of a parent cluster, with
negligible tidal effects,  could still produce a significant warp in
the outer part of a galactic disk.
We present results of numerical calculations for simplified models that show,
 for instance, that a satellite with the (baryonic) mass and distance
 of the Magellanic clouds can distort the axisymmetric field of the Milky Way
enough to produce a warp of the magnitude (and position) observed.
Details of the warp geometry remain to be explained: we use a static
configuration that can produce only warps
with a straight line of nodes. In more realistic simulations one must
reckon with the motion of the perturbing body, which sometimes occurs
on time scales not much longer than the response time of the disk.
\end{abstract}

\section{introduction}

It is not yet known what exactly induces and maintains galactic warps.
 It may well be that a number of mechanisms--from among those proposed already, or some
 new ones--act
together or alone to produce this ubiquitous phenomenon. Some proposed mechanisms rely on 
a dark galactic halo as a direct actuator or as a mediator of perturbations
(for reviews with extensive references see \cite{briggs90},
\cite{binney92}, and \cite{bmer}).
The modified dynamics (MOND) repudiates dark halos, but it offers a new mechanism
that increases the warping efficacy of external perturbers over and above their possible
 tidal effects, which, notoriously, are too weak. This results from the nonlinearity of MOND
 and is most clearly demonstrated in the case where a system (a galaxy) falls
 in an
 external field that by itself is approximately constant in space.
In a linear theory, such as Newtonian gravity, the constant external field has no effect
on the internal dynamics of the system (motions with respect to its center of mass);   
in MOND it very much does.
 When the external field dominates the internal field of the system it is
 easy to deduce what its effects are, as discussed  e.g. in \cite{mgbk},
 and \cite{mil86}. In the present context the external acceleration is small compared with
 the internal ones at the position of the warp, which necessitates numerical
 studies.
\par
For our  mechanism to work in field galaxies,
 one or more perturbers must be present.
There is, indeed, growing evidence that the appearance of a warp in a galaxy is strongly 
correlated with the presence of nearby perturbers (see e.g. \cite{rc98}). Even galaxies that
 had been thought to be isolated might, in fact, not be so (\cite{shang}). Of course, perturber
 companions have always been suspected, but their direct tidal effects on disks seem to be too small.
\par
 The purpose of this letter is to demonstrate, by numerical solutions of simplified 
galaxy-perturber systems, that, with reasonable parameter values, this MOND effect  
produces galactic warps of the magnitude observed. We have not included effects
due to variations in the external field (due to the motion of the perturber, or to the
motion of the galaxy in a parent cluster). From the symmetry of our model problem
 the warps we produce have a straight line of nodes.

 The method is described in section 2; the results are detailed in sections 3 and 4;
conclusions are drawn in section 5.

\section{Method}

We use the nonrelativistic, modified-gravity formulation of MOND suggested by
\cite{mgbk}. The acceleration  field  $\vg=-\grad\phi$ produced by a mass distribution 
$\rho$ is derived from a potential $\phi$ that satisfies  
 \begin{equation} \label{mond_field_eq}\grad\cdot[\mu(|\grad\phi|/a_0) 
\grad\phi]= 4\pi G\rho \end{equation}   
 instead of the usual Poisson equation  $\grad\cdot \grad\phi=4\pi G\rho$,
  where  $\mu(x) \approx x$ for $x\ll 1$, and 
 $\mu(x)\approx 1$ for $x\gg 1$,
 and $a_0$ is the acceleration constant of MOND.     
The form  $ \mu(x)=x/\sqrt{1+x^2}$ has been used       
       in all rotation curve analyses, and we also use it here.
This nonlinear potential equation is solved numerically using multi-grid methods as detailed
in  \cite{thesis}, and adumbrated in \cite{stab}.
\par
We consider two classes of models. To simulate a far away companion, or the effect of the mean
 field of a cluster on a member galaxy, we solve for the field of a 
rigid disk in the presence of a given
external acceleration field $\vgx$. In this case the field equation
is solved subject to the boundary condition
 at infinity  $\phi_\infty(\vec r)=-\vec r\cdot\vgx$.
Then $\vgx$ is subtracted from $-\grad\phi$ to get the field
 relative to the galaxy. This latter determines the galaxy's internal dynamics,
warps, etc.
To simulate the effect of a nearby companion, exemplified here by
the effect of the Magellanic clouds (MC) on the Milky-Way (MW), we solve fully for a
 disk-plus-perturber system (in which case 
$\grad\phi\rightarrow 0$ at infinity). Then, the center-of-mass acceleration of the galaxy is 
computed using the surface-integral method given by eq.(14) of
 \cite{mgbk}, and is subtracted from the acceleration field to get the internal dynamics.
\par
In each case, after the acceleration field relative to the center of mass
of the galaxy is found, we find closed, nearly circular, nearly planar,
 test-particle orbits. The orbits are integrated for many periods to insure
 that, within our accuracy and patience, they are closed. Thus, inasmuch as 
adiabaticity is a good approximation, these are non-precessing orbits. 
They are also found to be stable under small changes in their initial
 conditions.   
These are taken to trace a warp, in the spirit of the tilted-ring model
(\cite{rogstad74}).

\section{An exponential disk in a constant external field}

We take the model galaxy to be an exponential disk smoothly truncated at a
 radius that we use as our unit length, $R_{cut}=1$, and
 with a scale length
 of $h=0.2$ in these units. The surface density is of the form $\Sigma_0
 \exp(-R/h)(1-R^4) $ for
$0 \leq R \leq 1$. The disk  lies in the $x-y$ plane. To optimize the warping effect we take the
external field to lie $45^o$ from the $x$ axis in the third quadrant
of the $x-z$ plane. Its absolute value is taken as $g_{ex}=0.01$ in units of $a_0$.
 (We work in units where $a_0=1$, $G=1$, so masses are given in units of 
$a_0 R^2_{cut}/G$.)
In a more extensive study we plan to calculate the effect as a function of the field direction
 (relative to the disk axis) and also to follow the test particle orbits as the external field
 changes with time to mimic the relative motion of the galaxy and perturber.
For the present, pilot study
 we ran models with two values of the disk mass
 $m=0.01$ and $m=0.04$. 
The (MOND) accelerations of the isolated disk models at $R=1$ are 
$\approx m^{1/2}= 0.1, 0.2$;
 i.e.  respectively, ten and twenty times larger than  
the external field. Both accelerations are small compared with 1 ($a_0$) so we are rather
deep in the MOND regime in the warp region. In the deep-MOND regime the theory has obvious scaling 
properties so the above parameters represent a family of models spanned e.g. by scaling by the same factor $m^{1/2}$
and $g_{ex}$, or $m$ and $R_{cut}^2$ (with $h/R_{cut}$ fixed).
\par
\begin{figure}[htp]
\centerline{ \psfig{figure=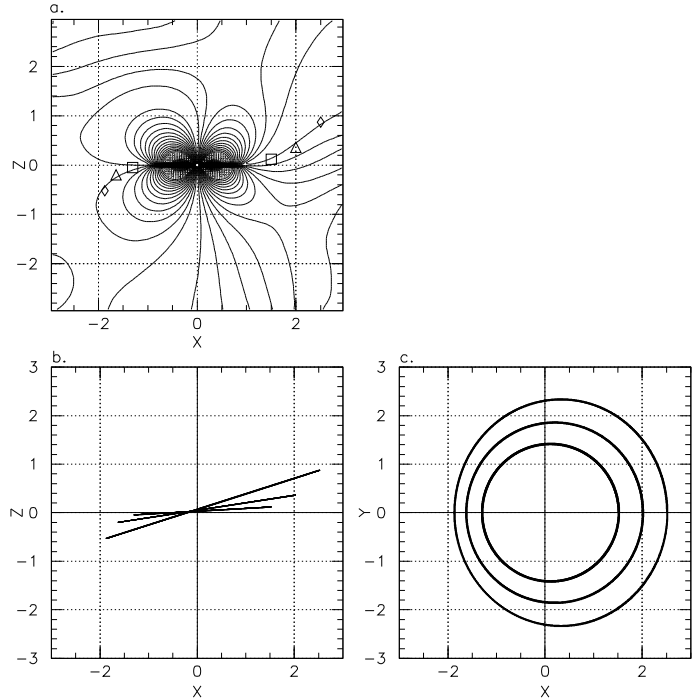,width=10cm}}
\caption{\protect\small The field geometry and some closed orbits for the $m=0.01$ case. The upper
 panel shows the lines of constant $|\vec r\times\grad\phi_c |$ in the $x-z$ plane. The lower panels show the 
projections of some closed orbits on the $x-z$ and $x-y$ planes.
The three symbols
$\Box$, $\triangle$, and $\diamond$ show their crossing points of the $x-z$
plane.}
\label{fig1}
\end{figure}
\begin{figure}[htp]
\centerline{ \psfig{figure=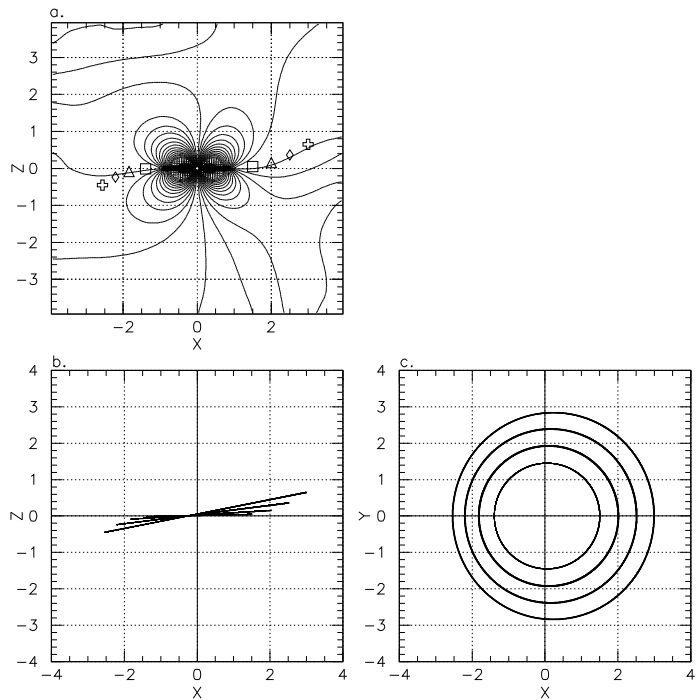,width=10cm}}
\caption{\protect\small The same as Figure 1 with $m=0.04$ for $4$ orbits.}
\label{fig2}
\end{figure}

The results are summarized
in Figures \ref{fig1} and \ref{fig2}.
We first show for each model a plot of the absolute value of the torque
$T\equiv |\vec r\times\grad\phi_c |$ in the $x-z$ plane containing the disk
 axis and the
 external field ($\grad\phi_c$ is the field in the center-of-mass frame).
 This is a quantity that brings out clearly the departure of 
the field from both axisymmetry and left-right symmetry.
 In the spherical case $T=0$ everywhere;
in the isolated-disk case the $T=0$ line is the $x$ axis (and the $z$ axis).
 This torque plot is also useful for homing in on closed orbits of the potential
whose center is near the galactic center, because these should cross the
$x-z$ plane near the zero-torque line. The orbits are found by actual 
 integration, starting from a set of initial conditions.
The projections of some such orbits are then shown.
  We surmise that in the spirit of the tilted-ring model they 
delineate the shape of the warp.

\section{A disk-plus-companion system--the effect of the Magellanic Clouds on the Milky Way.}
The disk of the Milky Way is known to be
warped beyond the solar circle (\cite{burke57,kerr57,henderson82}, and for a recent description
 and references \cite{bmer}). At galactic longitude $(l\approx 90^\circ)$
the HI disk curls steadily away from the plane.
At $(l\approx 270^\circ)$ the disk curves 
southward before turning back towards the plane (see \cite{bmer} for an 
analytic expression that approximates the warp shape beyond 11 kpc). 

The line of nodes in the tilted-ring picture is straight within the uncertainties, and is nearly perpendicular to the plane spanned by the inner-disk axis
and the radius vector to the Magellanic clouds. This makes the cloud system a
prime suspect in producing the warp.
It was, however, appreciated long ago that the tidal field of the Clouds, 
in their present position, is too small to
distort the disk to the extent observed. For example, \cite{hunter_toomre69}
estimated that a cloud mass
$\approx 10^{10} M_{\odot}$ would generate a warp of amplitude $\leq$ 70 pc
at a radius of 16 kpc.
It has, however, been suggested by \cite{weinberg95} that a ``live''
halo that actively responds to the perturbation of the clouds might augment the
effect to produce a warp of the observed magnitude and geometry.
\par
MOND, as we said, excludes a dynamically important halo, but might 
lead to a large enough warp due to the non-linear effect discussed above.
We model the system as follows. The MW is taken as a pure disk in the
 $x-y$ plane, centered at the origin, with
the cutoff, exponential surface-density law described in section 3
(with $h=0.2$); its
dimensionless mass is $M_{disk}=0.04$. The Magellanic clouds are represented by
one point mass $M_{sat}$ at a position whose dimensionless coordinates are
(2.52, 0, -1.63). This is at $15h$ from the center of the Galaxy, and
at the correct galactic latitude of the LMC. This ratio would correspond
for example to $h=3$ kpc (\cite{bmer}) and an LMC distance of 45 Mpc
 (\cite{mould}). The uncertainties in these parameters are still large.
Two mass ratios were considered: $M_{sat}/M_{disk}=0.1, 0.2$.
(The B luminosity ratio of the clouds to the MW is about 0.2. Since the
 baryonic $M/L$ values of the two might be different, reasonable values of
 the mass ratio lie between 0.1 and 0.4.) Other nearby galaxies are expected
to have a smaller effect than the LMC; for example, M31, despite its higher
mass, causes a rather smaller acceleration near the MW. 
\par
\begin{figure}[htp]
\centerline{ \psfig{figure=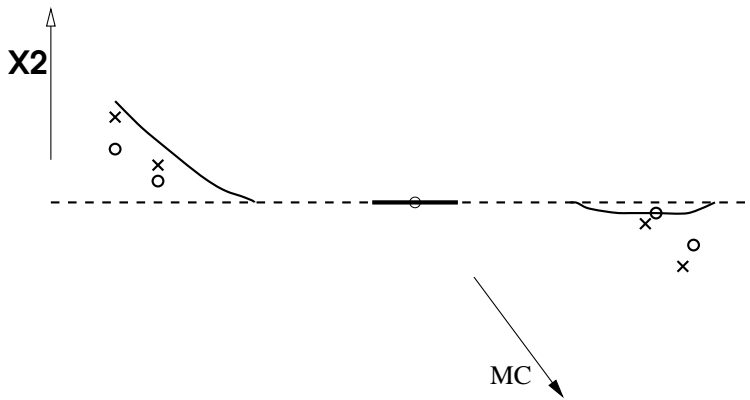,width=10cm}}
\caption{\protect\small  The calculated
 Milky Way warp. Shown are the crossing points
 of two stable, test-particle orbits for two satellite-to-disk mass ratios:
0.1 (circles), and 0.2 (crosses). The central bar marks one disk scale
 length (=3 kpc) on each side. The heavy line shows the observed  warp of
 the galaxy. The arrow marks the direction to the 
perturber representing the MC. The vertical scale is stretched by a factor of
 2 relative to the horizontal scale.}
\label{fig3}
\end{figure}

In Figure \ref{fig3} we show, for each mass ratio,
 two closed, quasi-circular, stable orbits beyond the cutoff radius of 
the disk--presumed to delineate our calculated warp--together with a
 representation of the observed warp (as given by the formula in \cite{bmer}).

\section{Conclusions and discussion}
We see that for a constant external field  
whose ratio to the field of the isolated disk at $5h$ is only (5-10)$\%$,
a noticeable warp beyond this radius is indicated by test-particle orbits.
This acceleration ratio increases in proportion to the galactocentric radius,
 but, still, rotational velocities will be affected only 
little even up to $(10-15)h$.
 Note that the warp is not symmetric but is less pronounced on 
the attracting side of the field.
\par
We used an external-field inclination that is favorable for an S-shape warp.
When the field is in the disk plane, the axisymmetry is broken but the 
up-down one is not, so no warp will be induced. If the field is perpendicular to
 the disk, axisymmetry is preserved but not the up-down one; this might 
induce bowl-shaped warps. (In the limit of a highly dominant, perpendicular,
 external field, the analytic results in \cite{mil86} show that
the geometry remains up-down symmetric, but for a weak perturbing field
this is not so.)
\par
Regarding the results for the MW-MC system, we see that even for a mass
 ratio of 0.1 a satellite at the position of the MC produces enough field
 distortion to accommodate inclined, quasi-circular orbit that rise to $0.25h$
at $R=6h$ on one side, and to a height of $0.2h$ at $R=5.5h$ on the other.
With a mass ratio of 0.2 the amplitude of the warp is close to that observed 
for the MW.
\par
Our analysis requires various improvements, which we hope to include in a 
future, more extensive analysis.
\par\noindent
1. A larger volume of the parameter space has to be surveyed. This includes more
values of the relative strength of the perturbation, different disk-perturber 
alignments (leading perhaps to a wider varieties of warp shapes), more complex
perturbations such as two or more satellites, which would bend
the
line of nodes at larger radii, where we cannot approximate the combined effect
 by a constant field. We would also have to study other galaxy mass
 distributions. For example, we expect that if a considerable fraction of the 
galaxy mass is put in a round bulge, a warp will form more easily.
For the same reason, if $h$ is smaller (but the MC distance remains the same)
the warp will be stronger at the same position.
\par\noindent
2. Viewing the warp as an envelope of test-particle orbits may be too naive.
 Certainly in more complicated geometries we expect orbits to cross and gas
dynamics must be considered.
\par\noindent
3. We must reckon with the fact that in many relevant cases the
geometry of the perturbation
 changes considerably during the response time of the disk
(say the orbital period at the position of the warp). This is true of
galaxies moving in or near the core of galaxy clusters; it is also
true for the MW-MC system. The warp geometry will thus not
just follow the perturbation adiabatically but, at larger radii, the geometry 
may reflect the past history of the perturbation (leading, among other things,
to curvature of the line of nodes).  
According to proper-motion observations and models of the MC and the Magellanic stream motion
(see e.g. \cite{lin_magellanic95}), the MC binary is moving on a nearly
polar orbit around the galaxy with a tangential velocity that is now comparable with the rotational velocity of the galaxy. This means that the radius
vector to the  MC changes its angle with the Galaxy's axis by $90^o$ during the
Galaxy's orbital period at about 15 kpc. This means that the adiabaticity
assumption we have made might be broken, and more and more so at larger radii.
Our subtraction of a constant center-of-mass acceleration is then also not 
valid.
This could lead to a more complicated warp geometry than the integral-sign
shape that we get with adiabaticity. Because the MC orbit is nearly polar
we expect the line of nodes to remain straight. 
\par\noindent
4. We think self-gravity of the mass in the warped region is not so important.
This is because the relative contribution of the warped mass to the
 acceleration field is small everywhere, even within the warp itself.
(The surface density in the warp is smaller than the integrated surface
density there.) This contribution can then be treated as a
 perturbation--linearizing in it the MOND field equation--and in MOND such
 density perturbations produce an even weaker effect than in
 Newtonian dynamics
(hence the added stability in MOND). So we can at least expect that 
the nonlinearity of MOND will not beget some peculiar amplification of self
 gravity. But we do not really know what these effects might be--a point that
 has to be checked numerically.
\par
We plan to perform $(N+1)$-body simulations whereby the $N$-body warped
 disk and
the point-mass perturber orbit each other.
This will account for non-adiabaticity and for self gravity in the disk, and
also partly for point 2 above. 
\par
We thank James Binney for helpful suggestions and for comments on the 
manuscript, and the referee for improving suggestions.



\begin{thebibliography}{}
\bibitem[Bekenstein $\&$ Milgrom (1984)]{mgbk}{Bekenstein. J., $\&$ Milgrom, M. 
1984,  ApJ, 286, 7}
\bibitem[Binney 1992]{binney92}{Binney, J.  1992,  ARA$\&$A, 30, 51}
\bibitem[Binney $\&$ Merrifield 1998]{bmer}{ Binney, J., $\&$ Merrifield, M.
 1998, Galactic Astronomy, (Princeton, Priceton Univ. Press)}

\bibitem[Brada (1996)]{thesis}{Brada, R. 1996, PhD thesis, Weizmann Institute}
\bibitem[Brada $\&$ Milgrom (1999)]{stab}{Brada, R., $\&$ Milgrom, M. 
1999,  ApJ, 519, 590}
\bibitem[Briggs 1990]{briggs90}{Briggs, F.H.  1990, ApJ, 352, 15}
\bibitem[Burke 1957]{burke57} {Burke, B.F. 1957,  AJ 62, 90}
\bibitem[Henderson, Jackson, $\&$ Kerr 1982]{henderson82}{Henderson, A.P. 
 Jackson, P.D., $\&$  Kerr, F.J. 1982, ApJ, 263, 116}
\bibitem[Hunter $\&$ Toomre  (1969)]{hunter_toomre69}{Hunter, C., $\&$ 
Toomre, A. 1969, ApJ, 155, 747}
\bibitem[Kerr  1957]{kerr57}{Kerr, F.J.  1957, AJ, 62, 93}
\bibitem[Lin, Jones,  $\&$ Klemola 1995]{lin_magellanic95}{Lin, D.N.C., 
Jones,B.F., $\&$  Klemola, A.R.  1995, ApJ, 439, 652}  
\bibitem[Milgrom (1986)]{mil86}{Milgrom, M. 1986,  ApJ, 302, 617}
\bibitem[Mould $\&$ al 1999]{mould}{Mould, J.R. $\&$ al 1999,
 preprint, astro-ph/9909260}
\bibitem[Reshetnikov $\&$ Combes 1998]{rc98}{ Reshetnikov, V., $\&$ Combes, F.
 1998, A$\&$A, 337, 9}
\bibitem[Rogstad, Lockhart, $\&$ Wright 1974]{rogstad74}{Rogstad, D.H.,
 Lockhart, I.A., $\&$ Wright, M.C.H.  1974,  ApJ, 193, 309}

\bibitem[Shang $\&$ al 1998]{shang}{Shang, Z. $\&$ al 1998, ApJL, 504, L23}
\bibitem[Weinberg  (1995)]{weinberg95}{Weinberg, M.D.  1995, ApJL, 455 L31}




\end{thebibliography}
\end{document}